\documentstyle[prl,aps,epsf,floats]{revtex}
\begin{document}
\draft
\twocolumn[\hsize\textwidth\columnwidth\hsize\csname @twocolumnfalse\endcsname
\title{Destruction of Strong Critical Spin Fluctuations by 
       Doping  \\
       in the 2D Hubbard Model
       }
\author{Bumsoo Kyung}
\address{Max Planck Institute for Physics of Complex Systems, 
         Noethnitzer Str. 38, 01187 Dresden, Germany} 
\date{23 January 1998}
\maketitle
\begin{abstract}

   We present the doping dependence of the spectral functions,
density of states and low frequency behavior of the self-energy
for the 2D Hubbard model  
on the basis of our recently developed theory for the Hubbard model.
Strong 2D critical spin fluctuations dominating near half-filling
are completely destroyed
by 13 and 20 $\%$ doping concentrations for $U=4$ and 8, 
respectively.
Below these concentrations the imaginary part of the 
self-energy vanishes quadratically in frequency near the Fermi energy,  
a characteristic feature for the Fermi liquid.
\end{abstract}
\vskip2pc]
\narrowtext

   High T$_{c}$ superconductors found in cuprates have 
exhibited a rich phase diagram in doping ($x=1-n$) and temperature ($T$)
plane\cite{Almasan:1991}.
At half-filling the materials are antiferromagnetic insulators
over a wide range of temperature up to 200-300 K, and 
away from half-filling
antiferromagnetic long-range order is destroyed 
by 2-3 $\%$ doping and further doping makes the cuprates be
metallic and eventually superconducting.
Right after the discovery of high T$_{c}$ superconductors,
the Hubbard model was proposed by Anderson\cite{Anderson:1987}
as the simplest model which might be able to describe
the rich and anomalous phase diagram in the copper oxides.
Although it was solved exactly in one-dimension\cite{Lieb:1968},
the exact solution in more than 2-dimensions is not known yet. 
Nontherless, there is increasing numerical evidence%
\cite{White:1991,Dagotto:1994}
that for a half-filled 2D band, the $T=0$ ground state is  
antiferromagnetic insulating for all $U$, while at finite temperatures
strong fluctuations destroy long-range 
antiferromagnetic order 
due to the Mermin-Wagner theorem\cite{Mermin:1966}.
Away from half-filling, however, there have been lack of 
reliable studies on the systematic change of dynamical 
properties by doping which include, particularly, 
the low frequency behavior 
of the self-energy.

   Although quantum Monte Carlo (QMC) calculations%
\cite{White:1991,Bulut:1994,Preuss:1995}
have exhibited
the systematic change of the  
spectral properties away from half-filling in the 2D Hubbard model, 
the lack of a reliable method for numerical analytical continuation
from QMC data containing inherent statistical errors
as well as finite size effect
have hindered a detailed study of some dynamical quantities such as the 
single particle self-energy.
While there have been extensive studies on 
the low frequency behavior 
of the self-energy
within 
fluctuation exchange (FLEX) approximation\cite{Dahm:1995},  
2D critical spin 
fluctuations are not properly taken into account in FLEX approximation. 
As a result, the scattering rates are linear (in frequency) over a 
wide range of frequency, but they always vanish quadratically near 
the Fermi energy 
for any available doping concentrations within FLEX.

   Recently we formulated a theory to the 2D 
Hubbard model in a manner free of finite 
size effect and numerical analytical continuation, yet 
containing the essential features of the 2D Hubbard model, i.e.,
the correct atomic limit at large $\omega$ and 2D spin fluctuations%
\cite{Kyung:19971}.
In this previous work for a half-filled 2D band,
2D critical spin fluctuations are found to give rise to a strong local maximum
at the Fermi energy  
in the scattering rates for 
low temperatures,
leading to a pseudogap in the spectral function. 
At zero temperature they are expected to yield eventually an antiferromagnetic 
insulating state.
Therefore, an important and natural question away from half-filling
is whether 
strong 2D critical spin fluctuations
are destroyed by doping or not, and if 
it is the case, how and at which doping concentration 
electrons lose their strong correlations, eventually leading 
to a Fermi liquid-like behavior.
In this Letter, we would like to answer this 
one of the central issues in the high T$_{c}$ superconductivity
for the first time.

   In order to take into account  
strong 2D critical fluctuations properly, 
we impose the following three exact sumrules to 
the spin, charge, and particle-particle susceptibilities%
\cite{Kyung:19971,Vilk:1997}:
\begin{eqnarray}
\frac{T}{N}\sum_{q}\chi_{sp}(q) & = & n-2\langle n_{\uparrow}n_{\downarrow}
                                         \rangle
                                                         \nonumber  \\
\frac{T}{N}\sum_{q}\chi_{ch}(q) & = & n+2\langle n_{\uparrow}n_{\downarrow}
                                         \rangle-n^{2}
                                                         \nonumber  \\
\frac{T}{N}\sum_{q}\chi_{pp}(q) & = & \langle n_{\uparrow}n_{\downarrow}
                                         \rangle
                                               \; .
                                                           \label{eq1}
\end{eqnarray}
$T$ and $N$ are the absolute temperature and 
number of lattice sites. 
$q$ is a compact notation for $(\vec{q},i\nu_{n})$ where
$i\nu_{n}$ are either Fermionic or 
Bosonic Matsubara frequencies. 
The dynamical spin, charge and particle-particle susceptibilities are 
calculated by 
\begin{eqnarray}
\chi_{sp}(q)&=&\frac{2\chi^{0}_{ph}(q)}{1-U_{sp}\chi^{0}_{ph}(q)}
                                                         \nonumber  \\
\chi_{ch}(q)&=&\frac{2\chi^{0}_{ph}(q)}{1+U_{ch}\chi^{0}_{ph}(q)}
                                                         \nonumber  \\
\chi_{pp}(q)&=&\frac{ \chi^{0}_{pp}(q)}{1+U_{pp}\chi^{0}_{pp}(q)}
                                               \; .
                                                           \label{eq2}
\end{eqnarray}
$\chi^{0}_{ph}(q)$ and 
$\chi^{0}_{pp}(q)$ are 
irreducible particle-hole and particle-particle susceptibilities, 
respectively, which are computed from
\begin{eqnarray}
\chi^{0}_{ph}(q) & = & - \frac{T}{N}\sum_{k}G^{0}(k-q)G^{0}(k)
                                                         \nonumber  \\
\chi^{0}_{pp}(q) & = &  \frac{T}{N}\sum_{k}G^{0}(q-k)G^{0}(k)
                                               \; ,
                                                           \label{eq3}
\end{eqnarray}
where 
$G^{0}(k)$ is the noninteracting Green's function.
$U_{sp}$, $U_{ch}$, and $U_{pp}$ in Eq.~\ref{eq2} are    
renormalized interaction constants for each channel which are 
calculated self-consistently 
by making an ansatz
$U_{sp} \equiv U\langle n_{\uparrow}n_{\downarrow} \rangle/
(\langle n_{\uparrow} \rangle
\langle n_{\downarrow} \rangle)$\cite{Vilk:1997}
in Eq.~\ref{eq1}.
By defining 
$U_{sp}$, $U_{ch}$, and $U_{pp}$   
this way, the Mermin-Wagner 
theorem as well as correct atomic limit for large $\omega$
are satisfied 
simultaneously\cite{Kyung:19971}. 
In order to find the chemical potential for interacting electrons, 
first
we calculate
Eqs.~(\ref{eq1})-(\ref{eq3}) and the self-energy 
(Eqs.~(\ref{eq1}) in Ref.~\onlinecite{Kyung:19971})
with the noninteracting Green's function whose 
noninteracting chemical potential gives 
a desired electron concentration. 
Then, 
the chemical potential for interacting electrons is determined 
in such a way that the calculated electron concentration 
with the interacting Green's function 
is the same as the desired value.
Throughout the calculations the unit of energy is $t$ and all energies 
are measured from the chemical potential $\mu$. 
We used a $128 \times 128$ lattice in momentum space 
and performed the calculations by means of 
well-established fast Fourier transforms
(FFT).
It should be also noted that we used a real frequency formulation
in Eqs.~(\ref{eq1})-(\ref{eq3}) to avoid any possible uncertainties
associated with numerical analytical continuation.

\begin{figure}
 \vbox to 7.5cm {\vss\hbox to -5.0cm
 {\hss\
       {\includegraphics{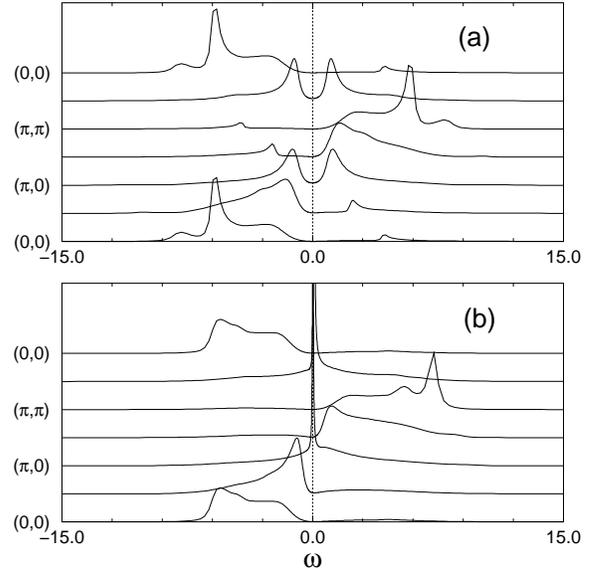}
       }
  \hss}
 }
\caption{Spectral functions for $U=4$ 
         (a) at $n=1.0$, $T=0.05$ and 
         (b) at $n=0.87$, $T=0.005$.}
\label{fig1}
\end{figure}

   We start in Fig.~\ref{fig1} by studying 
the single particle spectral functions for $U=4$
at $n=1.0$, $T=0.05$ (Fig.~\ref{fig1}(a)) and  
at $n=0.87$, $T=0.005$ (Fig.~\ref{fig1}(b)).  
Within our numerical calculations, we do not have the self-consistent 
solution for $U_{sp}$ at lower temperatures than 0.05 near half-filling.
Note that the spin-spin correlation length exponentially 
increases like
$\xi \sim exp(constant/T)$ at low temperatures near half-filling.
In Fig.~\ref{fig1}(a) for a half-filled band,  
an antiferromagnetic pseudogap is open at the Fermi energy along the 
antiferromagnetic zone boundary and a small shadow structure
is found in the opposite frequency side of  
the main peak  
at the other momenta.
At $n=0.87$ (Fig.~\ref{fig1}(b)), 
generally the spectral weight is transferred near the Fermi energy
and a shadow feature is strongly suppressed
throughout the Brillouin zone.
The spectral functions 
along the 
antiferromagnetic zone boundary
are so drastically changed
at this electron concentration
that 
the antiferromagnetic pseudogap is completely closed  
and a sharp quasiparticle peak 
begins to develop near the Fermi energy.

\begin{figure}
 \vbox to 7.5cm {\vss\hbox to -5.0cm
 {\hss\
       {\includegraphics{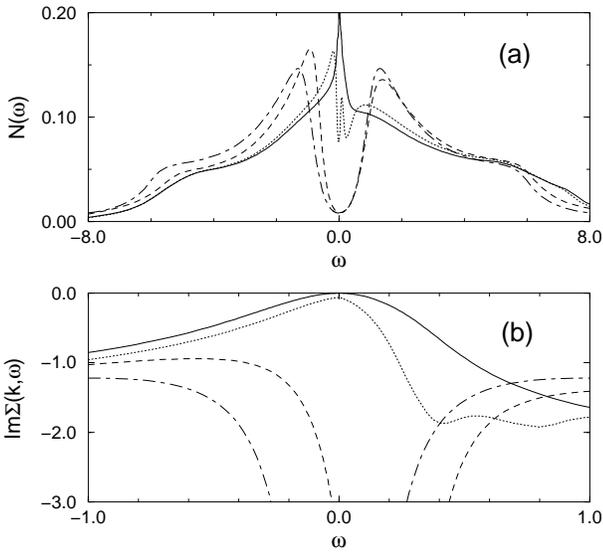}
       }
  \hss}
 }
\caption{(a) Density of states and 
         (b) low frequency domain of the imaginary part of the
         self-energy at the noninteracting Fermi surface for $U=4$ at
         different electron concentrations
         $n=1.0$, 0.95, 0.90, and 0.87 denoted as the dot-dashed, 
         dashed, dotted, and solid curves, respectively.
         For $n=1.0$ and 0.95, $T$ is 0.05, while
         for $n=0.90$ and 0.87, $T$ is 0.005.} 
\label{fig2}
\end{figure}

   In Fig.~\ref{fig2} we present the systematic change of the 
density of states and low frequency part of 
the self-energy 
by varying the electron concentration from 
$n=1.0$ to 0.95, 0.90 and 0.87.  
As the concentration is decreased, the spectral weight 
tends to 
move near the Fermi energy and to fill the antiferromagnetic pseudogap
(Fig.~\ref{fig2}(a)).
At $n=0.90$ (dotted curve), 
most of the antiferromagnetic pseudogap is closed and a small 
peak develops at the Fermi energy.
At $n=0.87$ (solid curve), 
the antiferromagnetic pseudogap is completely destroyed 
by doping with the appearance
of a narrow and sharp spectral weight
at the Fermi energy.
At this stage
a Fermi liquid-like 
behavior is expected to set in.
In Fig.~\ref{fig2}(b) the low frequency part of the scattering rates 
is shown for the same parameters as in 
Fig.~\ref{fig2}(a).  
At $n=1.0$ (dot-dashed curve), strong 2D critical spin fluctuations 
give rise to enormous 
scattering rates near the Fermi energy, eventually leading to
an antiferromagnetic insulator at zero temperature.
At $n=0.95$ (dashed curve), 
they are still persisting and shifted 
to the positive frequency side.
At $n=0.90$ (dotted curve), most of the critical fluctuations
are destroyed by doping with a small remnant in the positive 
frequency side.
At $n=0.87$ (solid curve), the 2D critical fluctuations
are completely washed out and the imaginary part of the self-energy 
vanishes quadratically in frequency near the Fermi energy, a characteristic 
feature for the Fermi liquid.

\begin{figure}
 \vbox to 7.5cm {\vss\hbox to -5.0cm
 {\hss\
       {\includegraphics{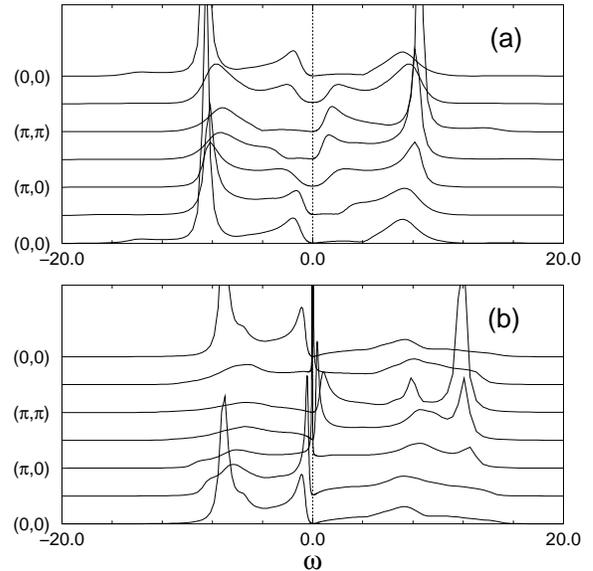}
       }
  \hss}
 }
\caption{Spectral functions for $U=8$ 
         (a) at $n=1.0$, $T=0.05$ and 
         (b) at $n=0.80$, $T=0.005$.}
\label{fig3}
\end{figure}

   In Fig.~\ref{fig3}  
the single-particle spectral functions for $U=8$
at $n=1.0$, $T=0.05$ (Fig.~\ref{fig3}(a)) and  
at $n=0.80$, $T=0.005$ (Fig.~\ref{fig3}(b)) are  
shown.
At $n=1.0$, four peaks are found in the spectral function, 
two peaks associated with the antiferromagnetic bands near the Fermi energy 
and the other two with the Hubbard bands in the intermediate frequency 
regime.
The Hubbard bands appear dispersionless since they arise from strong 
local repulsions, and for $U=8$ they are more pronounced 
than the antiferromagnetic bands%
\cite{Kyung:19971}.
The antiferromagnetic pseudogap is more visible along the 
antiferromagnetic zone boundary, but its total spectral weight
is much more suppressed compared with that for $U=4$
in Fig.~\ref{fig1}.  
At $n=0.80$ (Fig.~\ref{fig3}(b)), 
the spectral weight is transferred near the Fermi energy
throughout the Brillouin zone, as is the case for $U=4$
in Fig.~\ref{fig1}(b).
At this electron concentration,
the Hubbard bands are strongly suppressed  
and their significant spectral weight
moves near the Fermi energy to fill the antiferromagnetic pseudogap.
The sharp quasiparticle peaks 
appear near the Fermi energy.

\begin{figure}
 \vbox to 7.5cm {\vss\hbox to -5.0cm
 {\hss\
       {\includegraphics{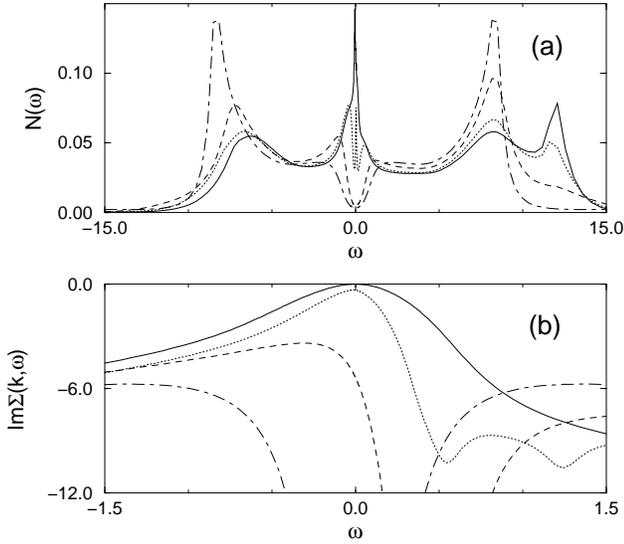}
       }
  \hss}
 }
\caption{(a) Density of states and 
         (b) low frequency domain of the imaginary part of the
         self-energy at the noninteracting Fermi surface 
         for $U=8$ at different electron concentrations
         $n=1.0$, 0.91, 0.85, and 0.80 denoted as the dot-dashed, 
         dashed, dotted, and solid curves, respectively.
         For $n=1.0$ and 0.91, $T$ is 0.05, while
         for $n=0.85$ and 0.80, $T$ is 0.005.} 
\label{fig4}
\end{figure}

   The density of states and low frequency part of  
the scattering rates
for $U=8$ at different electron concentrations from 
$n=1.0$ to 0.91, 0.85 and 0.80  
are presented
in Fig.~\ref{fig4}.   
As the concentration is decreased, a considerable spectral weight 
associated with the Hubbard bands
(Fig.~\ref{fig4}(a))
moves near the Fermi energy and starts to fill the antiferromagnetic pseudogap.
At $n=0.85$ (dotted curve), 
most of the antiferromagnetic pseudogap is closed and a small 
peak develops at the Fermi energy, as in Fig.~\ref{fig2}(a).
At $n=0.80$ (solid curve), 
the antiferromagnetic pseudogap is completely closed by doping and 
a narrow, sharp spectral weight appears at the Fermi energy.
In Fig.~\ref{fig4}(b) the low frequency part of the scattering rates
is also shown for the same parameters as in 
Fig.~\ref{fig4}(a).  
At $n=1.0$ and 0.91 (dot-dashed and dotted curves),  
strong 2D critical fluctuations 
are still dominating near the Fermi energy.
At $n=0.85$ (dotted curve), most of the 2D critical fluctuations
are already destroyed by doping with a small remnant.  
Eventually 20 $\%$ doping (solid curve) destroys
the 2D critical fluctuations
completely and the scattering rates 
approach zero quadratically in frequency near the Fermi energy, 
as in Fig.~\ref{fig2}(b)
for $U=4$.

\begin{figure}
 \vbox to 7.5cm {\vss\hbox to -5.0cm
 {\hss\
       {\includegraphics{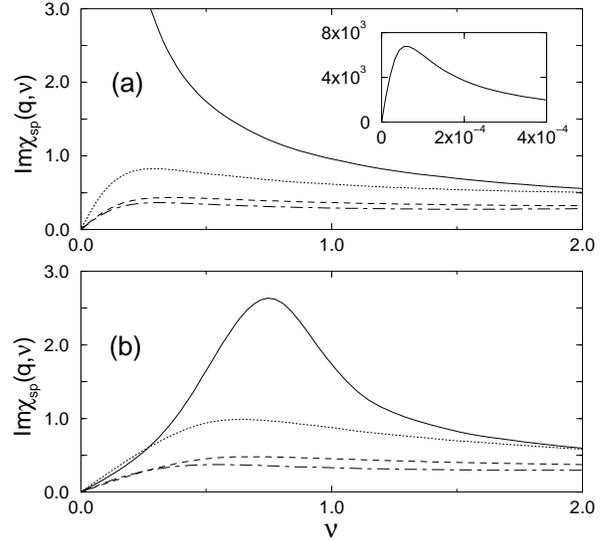}
       }
  \hss}
 }
\caption{Imaginary part of the  
         dynamical spin susceptibility for $U=8$ 
         (a) at $n=1.0$, $T=0.05$ and  
         (b) at $n=0.80$, $T=0.005$. 
         The dot-dashed, dashed, dotted and solid 
         curves denote $Im \chi_{sp}(\vec{q},\nu)$ 
         at $i=1$, 2, 3 and 4 
         for $\vec{q}=i(\pi/4,\pi/4)$, respectively.}
\label{fig5}
\end{figure}

   We plot in Fig.~\ref{fig5}
the imaginary parts of the dynamical spin susceptibility
for $U=8$
at $n=1.0$, $T=0.05$ (Fig.~\ref{fig5}(a)) and  
at $n=0.80$, $T=0.005$ (Fig.~\ref{fig5}(b)).
Both cases in Fig.~\ref{fig5}
show a similar behavior at small and intermediate momenta
except that the maxima for the former occur at a smaller 
energy scale than that for the latter, consistent with
the recent QMC calculations\cite{Preuss:1995}.
The most dramatic difference happens at $\vec{q}=(\pi,\pi)$
(solid curves).
For a half-filled 2D band (See the inset in Fig.~\ref{fig5}(a)), 
$Im \chi_{sp}(\vec{q},\nu)$  
shows a nearly divergent 
behavior at small frequencies 
(its maximum $6760/t$ occurs at $5.5 \times 10^{-5}t$), while
at $n=0.80$ 
(Fig.~\ref{fig5}(b))
$Im \chi_{sp}(\vec{q},\nu)$ exhibits only a broad maximum at 
$0.75t$.
This drastic change
in $Im \chi_{sp}(\vec{q},\nu)$  
is mainly responsible for the major differences 
discussed earlier 
for these two electron concentrations.

   Before closing we comment on the low 
frequency behavior of the self-energy.
$Im \Sigma(\vec{k},\omega)$  
for $U=4$, $n=0.87$ 
and for $U=8$, $n=0.80$
are also examined on a much smaller energy scale from -0.1 to 0.1
which is not shown in this Letter.
The quadratic behavior is still found with a small constant shift
(due to a finite temperature effect).
For the former case, the scattering rates vanish as 
$\omega^{1.95}$ near the Fermi energy, while for the latter as
$\omega^{1.94}$. 
As is shown in Figs.~\ref{fig2} and \ref{fig4}, we do not also find any 
indication of a linear frequency dependence for the scattering rates.
Hence, as long as the 2D Hubbard model with only nearest neighbor
hopping is concerned, our calculations do not show 
the frequency dependence 
of the scattering rates from FLEX approximation%
\cite{Dahm:1995} or from 
the marginal Fermi liquid
hypothesis\cite{Varma:1989}.

   In summary,
the doping dependence of the spectral functions,
density of states and low frequency behavior of the self-energy
has been studied
for the 2D Hubbard model   
based on our recently developed theory for the Hubbard model.
Strong 2D critical spin fluctuations dominating near half-filling
are completely destroyed
by 13 and 20 $\%$ doping concentrations for $U=4$ and 8, 
respectively.
Below these concentrations the scattering rates 
vanish quadratically in frequency near the Fermi energy,  
a characteristic feature for the Fermi liquid.

    The author would like to thank Prof. P. Fulde,  
and Drs. S. Blawid, R. Bulla, T. Dahm,  
P. Kornilovitch, M. Laad, W. Stephan and numerous other colleagues 
in the Max Planck Institute
for Physics of Complex Systems 
for useful discussions.   
\end{document}